\documentclass{article}%
\usepackage{amsmath}
\usepackage{amsfonts}
\usepackage{amssymb}
\usepackage{graphicx}%
\begin{document}

\title{Scaling property of the statistical Two-Sample Energy Test}
\author{G. Zech\thanks{Universit\"{a}t Siegen, D-57068 Siegen, Germany, email:
zech@physik.uni-siegen.de}}
\maketitle

\begin{abstract}
The \emph{energy test} \cite{nim} is a powerful binning-free,
multi-dimensional and distribution-free tool that can be applied to compare a
measurement to a given prediction (goodness-of-fit) or to check whether two
data samples originate from the same population (two-sample test). In both
cases the distribution of the test statistic under the null hypothesis $H_{0}%
$, (correct prediction, same population) has to be obtained by simulation.
This poses computational problems if the data samples are large, but the
difficulty can be overcome with the help of a scaling property which relates
the distribution of small samples to the distribution of large samples.
Scaling has been made plausible in \cite{barter} by extensive simulations. In
this article an analytic proof is presented which makes the calculation of
$p$-values obtained by scaling more reliable.

\end{abstract}

\section{Introduction}

In the two-sample energy test \cite{nim}, two data samples of $n$ and $m$
elements characterized by the coordinates $\mathbf{x}_{i}$, $\mathbf{y}_{i}$,
respectively, in a real multi-dimensional space are compared. The test
statistic is
\begin{equation}
\phi_{nm}=\frac{1}{2n(n-1)}%
{\displaystyle\sum\limits_{i=1}^{n}}
{\displaystyle\sum\limits_{j\neq i}^{n}}
R(|\mathbf{x}_{i}-\mathbf{x}_{j}|)+\frac{1}{2m(m-1)}%
{\displaystyle\sum\limits_{i=1}^{m}}
{\displaystyle\sum\limits_{j\neq i}^{m}}
R(|\mathbf{y}_{i}-\mathbf{y}_{j}|)-\frac{1}{nm}%
{\displaystyle\sum\limits_{i=1}^{n}}
{\displaystyle\sum\limits_{,j=1}^{m}}
R(|\mathbf{x}_{i}-\mathbf{y}_{j}|)\;. \label{twosample}%
\end{equation}
Here $R(|\mathbf{x}_{i}-\mathbf{x}_{j}|)$ is a function of the distance $r$
between the elements $i$ and $j$. Consistency of the test has been proven for
$R$ being the normal, the logarithmic and the inverse power of the distance
and is likely to be valid also for other monotonically decreasing functions.
In the case of the three-dimensional $1/r$ distance function, the test
statistic $\phi$ can be interpreted as the electrostatic energy of a system of
$n$ negative charges of charge $1/n$ each and $m$ positive charges of charge
$1/m$ each. The expected energy takes its minimum if the distributions of the
negative and the positive charges are equal. This corresponds to the null
hypothesis $H_{0}$ and the distribution $f_{n,m}(\phi_{n,m})$. As the
components of $\phi$ are mean values of the identically distributed random
variables $x$, $y$, one might expect that the distributions of the components
of $\phi_{nm}$ would - with increasing $n,m$ - approach a normal distribution.
Due to the correlation of the elements, this is not the case and since the
analytic form of the distribution is not known, the distribution has to be
obtained by simulations. One has to represent $\phi_{nm}$ by the discrete
combination of the two samples: From the combined sample, two samples of sizes
$n$, $m$ are drawn randomly with replacement. For each pair, the statistic
$\phi$ is computed. This procedure is repeated many times to obtain the
distribution $\phi$. In cases where the samples contain a large number of
observations, the computing time for the construction of $\phi$ can become
excessively long. For a given size $\alpha$ of the test and a required maximal
uncertainty $\delta\alpha/\alpha$ a number of repetitions $k\geq
1/(\delta\alpha/\alpha^{2})$ is required. For $k$ combination, about
$k(n^{2}+m^{2}+nm)$ distance functions have to be computed. With, for
instance, $\alpha=0.001\,,\delta\alpha/\alpha=0.1\,,\ k\approx10^{4}$ and
$m,n$ of the order $10^{5}$ we arrive at \ about $10^{14}$ computations of
$R$. To reduce the computing time, in \cite{root} it is proposed to switch to
a histogram version of the energy test and a corresponding program has been
posted at the program package ROOT \cite{root1}.

In the following we describe the scaling method that relates the distribution
of large samples to that of small samples and avoids the loss of information
due to binning. In this way, the computing time is reduced by a large factor.

In the goodness-of-fit version of the energy test, a data sample is compared
to a theoretical prediction. The null hypothesis is then presented by a Monte
Carlo sample that is generated according to the prediction. The size of this
sample is chosen large compared to the data sample and similar computational
problems occur as in the two-sample case. In this situation it is recommended
to use the two-sample formalism with the scaling method. (In \cite{nim} the
hypothesis is put forward that the distribution of $\phi$ in the
goodness-of-fit problem might follow the general extreme value distribution
accompanied by a warning. In the mean time it has turned out that the
hypothesis is not valid.)

\section{The scaling hypothesis}

The scaling hypothesis relates the distribution $f$ of $\phi_{n,m}$ under
$H_{0}$ to that of $\phi_{sn,sm}$ where the sample sizes are increased by a
positive scaling factor $s$, i.e. $n,m\rightarrow sn,sm$. For $n,m\gg1$ the
distribution of $s\phi$ is independent of $s$: $f_{n,m}(\phi_{n,m}%
)=f_{n,m}(s\phi_{sn,sm})$. Scaling by $s$, the distribution shrinks by the
factor $s$ and, correspondingly, the $k$-th central moment $\mu_{k}$, if
different from zero, shrinks by $s^{k}$:%
\[
\mu_{k}(\phi_{sn,sm})=\mu_{k}(\phi_{n,m})/s\;.
\]
Fig. 1 shows distributions of the test statistics $40\phi_{40,40}$ and
$100\phi_{100,100}$ under $H_{0}$ for a logarithmic distance function and a
two-dimensional uniform distribution of of the observations. Even though the
numbers are low, scaling is realized quite well except for very small values
of $\phi$.%
\begin{figure}
[ptb]
\begin{center}
\includegraphics[
height=3.6837in,
width=4.7015in
]%
{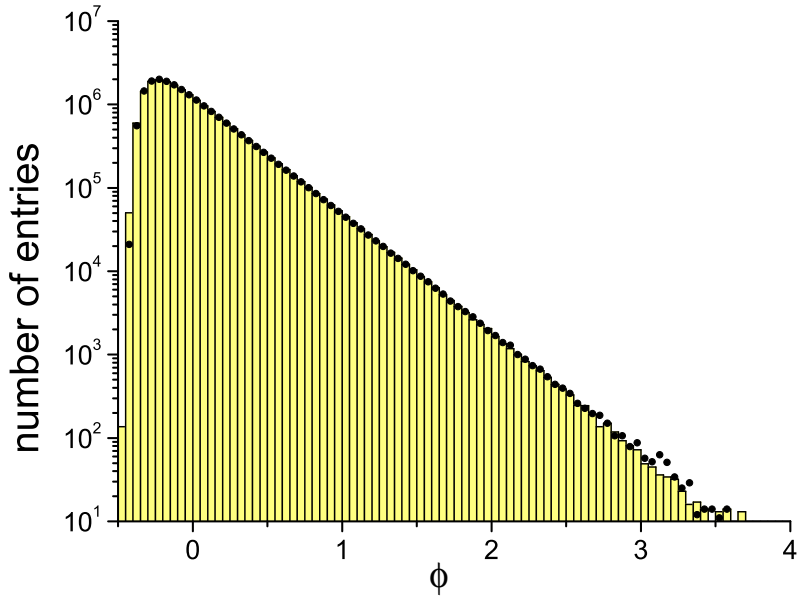}%
\caption{Distribution of the test statistic $n\phi_{nn}$ for $n=40$
(histogram) and $n=100$ (dots). }%
\label{scalingeps}%
\end{center}
\end{figure}

To demonstrate the scaling of $\phi$, it is sufficient to demonstrate the
scaling of the moments. As the expected value $<\phi>$ of $\phi$ is zero, the
central moments correspond to the expected values of powers of $\phi$,
$\mu_{k}=<\phi^{k}>$.

\section{Proof of the scaling property}

To proceed, we modify the notation. We abbreviate the three contributions of
$\phi_{nm}$ by $\phi_{1}$, $\phi_{2}$ and $\phi_{3}$. We set $<\phi_{1}%
>=E_{1}$, $<\phi_{2}>=E_{2}$, $<\phi_{3}>=E_{3}$ and $\phi_{i}=E_{i}+e_{i}$
and have $E_{1}+E_{2}+E_{3}=0$, because the expected value of the total energy
is equal to zero under $H_{0}$. We remain with $\phi=e_{1}+e_{2}+e_{3}$ and
$<e_{i}>=0$. The expected value of the distance function be $<R(|\mathbf{x}%
_{i}-\mathbf{x}_{j}|)>=a$, and $R(|\mathbf{x}_{i}-\mathbf{x}_{j}%
|)=a+\varepsilon_{ij}$, with $\varepsilon_{ii}=0$ and $<\varepsilon_{ij}>=0$.
The relation (\ref{twosample}) now reads
\[
\phi_{nm}=\frac{1}{2n(n-1)}%
{\displaystyle\sum\limits_{i=1}^{n}}
{\displaystyle\sum\limits_{j\neq i}^{n}}
\varepsilon_{ij}^{(1)}+\frac{1}{2m(m-1)}%
{\displaystyle\sum\limits_{i=1}^{m}}
{\displaystyle\sum\limits_{j\neq i}^{m}}
\varepsilon_{ij}^{(2)}-\frac{1}{nm}%
{\displaystyle\sum\limits_{i=1}^{n}}
{\displaystyle\sum\limits_{,j=1}^{m}}
\varepsilon_{ij}^{(3)}\;.
\]
We set further $C_{2}=<\varepsilon_{ij}^{2}>$ for $i\neq j$. The random
variables $\varepsilon_{ij}$ and $\varepsilon_{ik}$ of the same set are
correlated. The correlation is set $C_{1}=<\varepsilon_{ij}\varepsilon_{ik}>$
for $j\neq i\neq k$. There exist also higher order correlations like
$<\varepsilon_{ij}\varepsilon_{ik}\varepsilon_{il}>$ or $<\varepsilon
_{ij}\varepsilon_{ik}\varepsilon_{kl}\varepsilon_{ol}>$ and similar for higher
moments. Expected values like $<\varepsilon_{ij}\varepsilon_{lk}>$ and
$<\varepsilon_{ij}\varepsilon_{ik}\varepsilon_{ol}>$ are zero, because not all
random variables have a common index with another random variable.

\subsection{The second moment}

We compute the second moment $\mu_{2}$ explicitly.%
\begin{align}
\mu_{2}  &  =<\phi_{nm}^{2}>\nonumber\\
&  =<(e_{1}+e_{2}+e_{3})^{2}>\nonumber\\
&  =<e_{1}^{2}>+<e_{2}^{2}>+<e_{3}^{2}>+2(<e_{1}e_{3}>+<e_{2}e_{3}>)\;.
\label{e123}%
\end{align}
The parameters $e_{1}$ and $e_{2}$ are not correlated and $<e_{1}e_{2}>=0$.
The first term $<e_{1}^{2}>$ is
\begin{equation}
<e_{1}^{2}>=\frac{1}{(2n(n-1))^{2}}%
{\displaystyle\sum\limits_{i,j,k,l=1}^{n}}
<\varepsilon_{ij}\varepsilon_{kl}>\;. \label{mu4cor}%
\end{equation}
The two random values are independent if all four indices are different and
the expected value is zero. We remain with%
\begin{align*}
&  <e_{1}^{2}>=\frac{1}{(2n(n-1))^{2}}%
{\displaystyle\sum\limits_{i,j,k=1}^{n}}
(<\varepsilon_{ij}\varepsilon_{ik}>+<\varepsilon_{ij}\varepsilon
_{ki}>+<\varepsilon_{ji}\varepsilon_{ik}>+<\varepsilon_{ji}\varepsilon
_{ki}>)\;,\\
&  <e_{1}^{2}>=\frac{4}{(2n(n-1))^{2}}%
{\displaystyle\sum\limits_{i,j,k=1}^{n}}
<\varepsilon_{ij}\varepsilon_{ik}>\;.
\end{align*}
We have used that the four terms in the first line are equal. We have $n$
choices for $i$, $n-1$ for $j$ and $n-2$ for $k$ if all three indices are
different. The remaining terms of (\ref{e123}) behave similarly. The results
are summarized in Table \ref{tabmu2} were also the asymptotic values for
$n\gg1$ are given.

\begin{table}[ptb]
\caption{Contributions to $\mu_{2}$.}%
\label{tabmu2}
\centering
\begin{tabular}
[c]{|l|l|l|l|l|}\hline
term & norm & \# combinations $C_{1}$ & \# comb. $C_{2}$ & product\\\hline
$<e_{1}^{2}>$ & $1/(2n(n-1))^{2}$ & $4n(n-1)(n-2)$ & $2n(n-1)$ &
$\frac{2(n-2)C_{1}+C_{2}}{2n(n-1)}\approx C_{1}/n$\\
$<e_{2}^{2}>$ & $1/(2m(m-1))^{2}$ & $4m(m-1)(m-2)$ & $2m(m-1)$ &
$\frac{2(m-2)C_{1}+C_{2}}{2m(m-1)}\approx C_{1}/m$\\
$<e_{3}^{2}>$ & $1/(nm)^{2}$ & $nm(n+m-2)$ & $nm$ & $\frac{(n+m-2)C_{1}+C_{2}%
}{nm}\approx C_{1}\frac{n+m}{nm}$\\
$2<e_{1}e_{3}>$ & $-2/(2n^{2}m(n-1))$ & $2nm(n-1)$ & $0$ & $-\frac{2C_{1}}{n}
$\\
$2<e_{2}e_{3}>$ & $-2/(2nm^{2}(m-1))$ & $2nm(m-1)$ & $0$ & $-\frac{2C_{1}}{m}
$\\\hline
\end{tabular}
\end{table}

Adding all contributions we obtain:%
\begin{align}
\mu_{2}  &  =\frac{2(n-2)C_{1}+C_{2}}{2n(n-1)}+\frac{2(m-2)C_{1}+C_{2}%
}{2m(m-1)}+\frac{(n+m-2)C_{1}+C_{2}}{nm}-\frac{2C_{1}}{n}-\frac{2C_{1}}%
{m}\label{mu2}\\
&  =\frac{[(m+n)^{2}-3(n+m)+2](C_{2}/2-C_{1})}{nm(n-1)(m-1)}\approx
\frac{(m+n)^{2}}{n^{2}m^{2}}(C_{2}/2-C_{1})\;.\nonumber
\end{align}

\bigskip In the special case $n=m$ we obtain:%

\[
\mu_{2}=\frac{[2n^{2}-3n+1](C_{2}-2C_{1})}{n^{2}(n-1)^{2}}\approx\frac
{2(C_{2}-2C_{1})}{n^{2}}\;.
\]

We realize that scaling $n,m$ by $s$ reduces $\mu_{2}$ by $s^{2}$. This factor
is due to the correlation requirement which reduces the summing indices in
(\ref{mu4cor}) from $4$ to $3$ and to the fact that the leading terms in $n$
and $m$ cancel in the numerator.

The precision of the scaling approximation for finite $n$ can be estimated
from the expressions in the last line: As an example with $n=m=100$ we find
that the approximation underestimates the standard deviation of the
distribution by a factor of $1.0025$, very close to one and independent of the
distance function.

\subsection{Higher moments}

The explicit calculation of the higher moments is tedious. We have to be
satisfied with the proof that the moments scale for large $n,m$ like $\mu
_{k}(sn,sm)=\mu_{k}(n,m)/s^{k}$. For $n=m$ this means that $\mu_{k}/n^{k}$ is
independent of $n$. To reduce the number of terms we change the notation to%
\[
\phi_{nm}=%
{\displaystyle\sum\limits_{i,j=1}^{n+m}}
c_{ij}\varepsilon_{ij}
\]
and evaluate $\mu_{k}=<\phi^{k}>$ where the weights $c_{ij}$, their asymptotic
values and the asymptotic number of contributions are given in Table
\ref{tabfilow}.

First we consider $\mu_{3}$.%
\begin{align*}
\mu_{3}  &  =<\left[
{\displaystyle\sum\limits_{i,j=1}^{n+m}}
c_{ij}\varepsilon_{ij}\right]  ^{3}>\\
&  =%
{\displaystyle\sum\limits_{i,j,k,l,o,p=1}^{2n}}
c_{ij}c_{kl}c_{op}<\varepsilon_{ij}\varepsilon_{kl}\varepsilon_{op}>\;.
\end{align*}
Only terms where all three factors are correlated, like $<\varepsilon
_{ij}\varepsilon_{il}\varepsilon_{ip}>$, $<\varepsilon_{ij}\varepsilon
_{il}\varepsilon_{lp}>$ etc. survive and the order of the sum is reduced by
two:%
\begin{equation}
\mu_{3}=%
{\displaystyle\sum\limits_{i,j,l,p=1}^{n+m}}
c_{ij}c_{il}c_{lp}<\varepsilon_{ij}\varepsilon_{il}\varepsilon_{lp}>+\text{
}permutations\;. \label{mu3}%
\end{equation}
Again the leading terms should cancel and $\mu_{3}\sim O(1/(n^{3-p}m^{p})$, is
of "third order" in the denominator. To show this, we selected an arbitrary
term of the sum,

\begin{table}[ptb]
\caption{Contributions to $\phi$ in leading order}%
\label{tabfilow}
\centering
\begin{tabular}
[c]{|l|l|l|l|l|}\hline
$i$ & $j$ & $c_{ij}$ & ${\displaystyle\sum\limits_{i,j}1}$ & product, 1-st
O.\\\hline
$i\leq n$ & $j\leq n$ & $1/(2n(n-1))\approx1/(2n^{2})$ & $\approx n^{2}$ &
$1/2$\\
$i\leq n$ & $j>n$ & $-1/(2nm)$ & $\approx nm$ & $-1/2$\\
$i>n$ & $j\leq n$ & $-1/(2nm)$ & $\approx nm$ & $-1/2$\\
$i>n$ & $j>n$ & $1/(2m(m-1))\approx1/(2m^{2})$ & $\approx m^{2}$ &
$1/2$\\\hline
\end{tabular}
\end{table}%

\[
T=C_{3}%
{\displaystyle\sum\limits_{i,j,l,p=1}^{n+m}}
c_{ij}c_{il}c_{lp}\;,
\]
with $C_{3}=<\varepsilon_{ij}\varepsilon_{il}\varepsilon_{lp}>$ and use the
right hand results of Table \ref{tabfilow}.

For $i\leq n$ summing over $j$, we get for $\ c_{ij}$ and $n,m\gg1$ $n$ times
$1/(2n)$and $m$ times $-1/(2m)$, such that the sum becomes zero. The
contribution of the leading terms vanishes. For $i>n$ only the signs are
exchanged. Summing over an arbitrary index that occurs only once the leading
terms cancel such that the order of the numerator is reduced by one unit.

The corresponding result is obtained for the higher moments. For the moment
$\mu_{k}$ the denominator is of order $2k$ from the $k$ normalization factors.
Initially also the numerator is of order $2k$. We have $2k$ indices from which
$k$ have to be identical. The remaining $k+1$ indices lead to a numerator of
order $k+1$. The sum over an arbitrary index that occurs only once, the
leading terms in $n,m$ cancel such that the order is again reduced by one.
Consequently $\mu_{k}$ is of order $k/2k=1/k$ as required by the scaling
hypothesis. The exact form of the moments is complicated, but has to be
symmetric in $n$ and $m$.

\subsection{Conclusion}

The distribution of the test statistic $\phi_{nm}$ under $H_{0}$ for large
$n,m$ can asymptotically, $n,m\gg1$, be obtained by scaling the distribution
for moderate observation numbers $n/s$, $m/s$ with $s$ a scaling factor. The
scaling hypothesis is independent of the form of the distance function $R$ and
therefore also independent of the dimension of the sample space.

\end{document}